\documentclass[aps,prd,showpacs,floatfix,nofootinbib,twocolumn,10pt]{revtex4}

\usepackage{color,amsmath,amssymb,graphicx,latexsym,subfigure}
\usepackage{threeparttable,multirow,txfonts}

\def\aap{Astron.\ Astrophys.\ }
\def\apj{Astrophys.\ J.\ }
\def\apjl{Astrophys.\ J.\ Lett.\ }

\def\aj{Astron.\ J.\ }

\def\prd{Phys.\ Rev.\ D\ }
\def\prl{Phys.\ Rev.\ Lett.\ }

\def\jcap{J.\ Cosmol.\ Astropart.\ Phys.\ }

\begin{document}

\title{Nearby source interpretation of differences among light and medium
composition spectra in cosmic rays}

\author{Qiang Yuan$^{a,b,c}$\footnote{yuanq@pmo.ac.cn}}
\author{Bing-Qiang Qiao$^{a,d}$}
\author{Yi-Qing Guo$^{d}$}
\author{Yi-Zhong Fan$^{a,b}$}
\author{Xiao-Jun Bi$^{d,e}$}

\affiliation{
$^a$Key Laboratory of Dark Matter and Space Astronomy, Purple Mountain
Observatory, Chinese Academy of Sciences, Nanjing 210008, P.R.China \\
$^b$School of Astronomy and Space Science, University of Science and
Technology of China, Hefei 230026, P.R.China\\
$^c$Center for High Energy Physics, Peking University, Beijing 100871,
P.R.China\\
$^d$Key Laboratory of Particle Astrophysics, Institute of High Energy
Physics, Chinese Academy of Sciences, Beijing 100049, China\\
$^e$University of Chinese Academy of Sciences, Beijing 100049, China
}

\begin{abstract}
Recently the AMS-02 reported the precise measurements of the energy
spectra of medium-mass compositions (Neon, Magnesium, Silicon) of
primary cosmic rays, which reveal different properties from those of
light compositions (Helium, Carbon, Oxygen). Here we propose a nearby
source scenario, together with the background source contribution,
to explain the newly measured spectra of cosmic ray Ne, Mg, Si, and
particularly their differences from that of He, C, O. Their differences
at high energies can be naturally accounted for by the element abundance
of the nearby source. Specifically, the abundance ratio of the nearby 
source to the background of the Ne, Mg, Si elements is lower by a 
factor of $\sim1.7$ than that of the He, C, O elements. 
Such a difference could be due to the abundance difference of the
stellar evolution of the progenitor star or the acceleration 
process/environment, of the nearby source.
This scenario can simultaneously explain the high-energy spectral
softening features of cosmic ray spectra revealed recently by
CREAM/NUCLEON/DAMPE, as well as the energy-dependent behaviors of
the large-scale anisotropies. It is predicted that the dipole
anisotropy amplitudes below PeV energies of the Ne, Mg, Si group 
are smaller than that of the He, C, O group, which can be tested 
with future measurements.
\end{abstract}

\date{\today}

\pacs{95.85.Ry,96.50.S-,98.38.−j,94.20.wc}

\maketitle

\section{Introduction}

The measurements of the energy spectra of Galactic cosmic rays (CRs)
have entered a precise era, thanks to the contributions of a series of
new experiments such as PAMELA, AMS-02, CALET, and DAMPE. Several new
features of the CR spectra have been revealed recently, including the
hundred-GV hardenings \cite{2009BRASP..73..564P,2010ApJ...714L..89A,
2011Sci...332...69A,2015PhRvL.114q1103A,2015PhRvL.115u1101A,
2017PhRvL.119y1101A,2019PhRvL.122r1102A,2019SciA....5.3793A}
and $\sim10$ TV softenings \cite{2017ApJ...839....5Y,
2018JETPL.108....5A,2019SciA....5.3793A}. These new results challenge
our traditional understanding about the framework of CR production
and propagation, imposing new processes or ingredients of the CR
problems (e.g., \cite{2011ApJ...729L..13O,2011PhRvD..84d3002Y,
2012ApJ...752...68V,2012ApJ...752L..13T,2012PhRvL.109f1101B,
2016ApJ...819...54G,2018PhRvD..97f3008G,2019arXiv190705987K,
2020FrPhy..1524601Y,2020arXiv200313635F}).

Very recently, the AMS-02 group reported the measurements of the
primary CR spectra of mdeium-mass compositions, including the Neon
(Ne), Magnesium (Mg), and Silicon (Si) \cite{2020PhRvL.124u1102A}.
Spectral hardenings above $\sim200$ GV have been clearly revealed,
consistent with those of other nuclei. Unexpectedly, the rigidity
dependence of the mdeium-mass group shows distinct properties from
that of lighter compositions above 86.5 GV, which is supposed to be
an indication of two different classes of primary CRs
\cite{2020PhRvL.124u1102A}.

A natural explanation of the AMS-02 results would be a background
plus nearby source model, in which the nearby source contributes a
small fraction of the CR fluxes above a few hundred GV of rigidities
\cite{2013APh....50...33S,2015ApJ...809L..23S,2019JCAP...10..010L,
2019JCAP...12..007Q}. This model was shown to be able to explain
also the softening behavior of the CR spectra above $\sim10$ TV
\cite{2017ApJ...839....5Y,2018JETPL.108....5A,2019SciA....5.3793A}.
Given proper direction of the nearby source (close to the birth place
of the Geminga supernova), the energy-dependences of the amplitudes and
phases of the large-scale anisotropies (e.g., \cite{1996ApJ...470..501A,
2006Sci...314..439A,2009ApJ...692L.130A,2016ApJ...826..220A,
2017ApJ...836..153A}) can be well recovered
\cite{2019JCAP...10..010L,2019JCAP...12..007Q}.
If the abundances of the medium-mass elements of the nearby source
are slightly lower than the average of background sources, the
resulting high-energy spectra of Ne, Mg, and Si would be softer
than that of lighter elements. If this scenario is correct, the CR
data provides very useful implications on the chemical composition
of the nearby source --- either its progenitor or the acceleration
process. This is a very important clue in identifying this nearby 
CR accelerator.

In this work, we work out this model in detail to fit the AMS-02
measurements. Compared with previous works Refs. 
\cite{2019JCAP...10..010L,2019JCAP...12..007Q}, we pay special attention
to the spectral differences between the He, C, O group and the Ne, Mg,
Si group as emphasized by the AMS-02 experiment. We argue that such
differences actually offer an additional support to the nearby source
model, and the precise measurements can help to infer the source
properties of CRs. In Sec. II we describe the framework and parameters 
of the model. In Sec. III we present the fitting results. We conclude our 
work in Sec. IV with some discussion of the properties of the nearby source.

\section{Model framework}

The sources of the model include two components, a background component
diffusively distributed in the Milky Way, and a nearby source. For the
background component, we adopt a broken power-law with an exponential 
cutoff form in rigidity to describe the injection spectrum. The break 
is to fit the low-energy spectra \cite{2019SCPMA..6249511Y}. For the 
nearby source component, a single power-law form with an exponential 
cutoff is assumed. The spatial distribution of the background source 
is parameterized as
\begin{equation}
f(r,z)=\left(\frac{r}{r_\odot}\right)^{\alpha}\exp\left[-\frac
{\beta(r-r_\odot)}{r_\odot}\right]\,\exp\left(-\frac{|z|}{z_s}\right),
\end{equation}
where $r_\odot=8.5$ kpc, $z_s=0.2$ kpc, $\alpha=1.69$, and $\beta=3.33$,
which roughly traces the distribution of supernova remnants
\cite{1998ApJ...504..761C} but slightly adjusted.

For the propagation of CRs in the Milky Way, we adopt a spatially-dependent
diffusion approach \cite{2012ApJ...752L..13T,2015PhRvD..92h1301T,
2016ApJ...819...54G,2016PhRvD..94l3007F}. Note that the original
motivation of the spatially-dependent diffusion was to explain the
hundred-GV spectral hardenings of CRs. In principle it is not necessary
to keep this requirement in our current model if we limit our studies
to the CR spectra only. Nevertheless, the spatially-dependent diffusion
assumption adopted here is well motivated by the HAWC observations of
extraordinary slow diffusion of particles around pulsars in the Galactic
plane \cite{2017Sci...358..911A} compared with that inferred from the
secondary CRs \cite{2017PhRvD..95h3007Y}, as well as the explanation of
the anisotropy amplitudes at very high energies ($>100$~TeV)
\cite{2019JCAP...10..010L,2019JCAP...12..007Q}.

The general picture is that CRs diffuse much slower in the Galactic
disk where many sources drive the medium to a very turbulent state,
and faster in the halo. The spatial diffusion coefficient $D_{xx}$
is parameterized as
\begin{equation}
D_{xx}(r,z,\rho) = F(r,z) D_0 \beta \left(\frac{\rho}{\rho_0}\right)^
{F(r,z)\delta_0},
\end{equation}
where $\beta$ is the particle velocity, $\rho$ is the rigidity, $D_0$
and $\delta_0$ is the normalization and power-law slope of the diffusion
coefficient in the halo (when $F(r,z)\to 1$). The spatially-dependent
part of the diffusion coefficient $F(r,z)$ is assumed to be inversely
correlated with the source distribution as
\begin{eqnarray}
F(r,z)&=&\frac{N_m}{1+f(r,z)} \nonumber\\
      &+& \left(1-\frac{N_m}{1+f(r,z)}\right)\cdot
      \min\left[\left ( \frac{z}{\xi z_h}\right )^n,1\right],
\end{eqnarray}
where $\xi z_h$ denotes the half thickness of the slow-diffusion halo,
$N_m$ is a normalization factor, and $n$ characterizes the sharpness
between the disk and halo. For $z\ll \xi z_h$ (the disk), the diffusion
coefficient is obviously anti-correlated with the source distribution
$f(r,z)$. The diffusion coefficient becomes to the traditional form
of $D_0\beta(\rho/\rho_0)^{\delta_0}$ in the halo. The reacceleration
effect can be characterized by a diffusion in the momentum space.
The momentum diffusion coefficient $D_{pp}$ relates to $D_{xx}$ via the
effective Alfvenic velocity $v_A$ \cite{1994ApJ...431..705S}, as
$D_{pp}D_{xx}=\frac{4p^2v_A^2}{3\delta(4-\delta^2)(4-\delta)}$,
where $\delta=F(r,z)\delta_0$.

In this work we use the DRAGON code \cite{2008JCAP...10..018E,
2017JCAP...02..015E} to calculate the propagation of CRs. The main
propagation parameters are: $D_0=4.87\times 10^{28}$~cm$^2$~s$^{-1}$,
$\delta_0=0.58$, $z_h=5.0$~kpc, $v_A=6.0$~km~s$^{-1}$, $N_m=0.62$,
$\xi=0.1$, and $n=4$.

\section{Results}

\begin{figure*}[!htb]
\includegraphics[width=\textwidth]{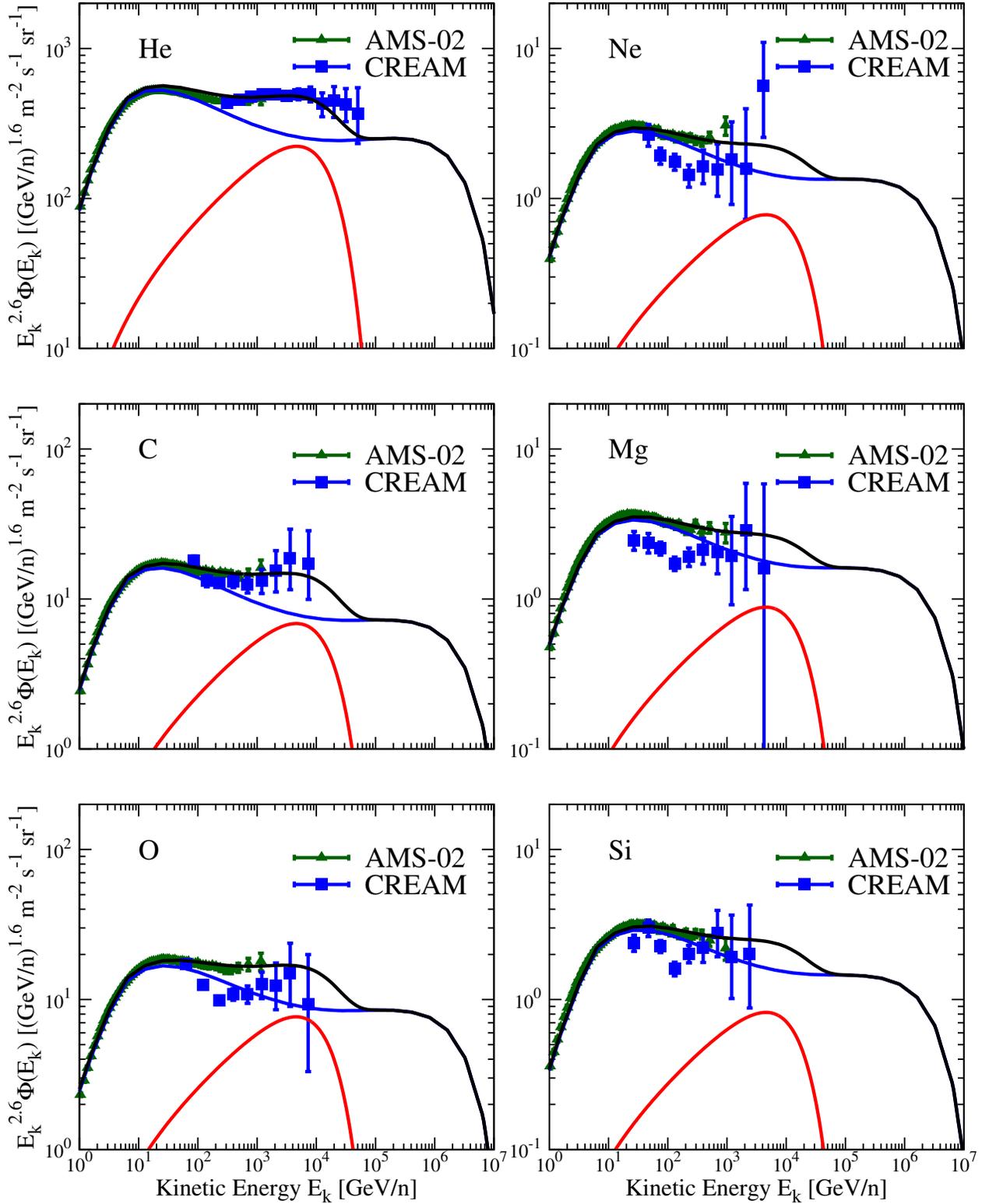}
\caption{Fittings of the energy spectra measured by AMS-02
\cite{2017PhRvL.119y1101A,2020PhRvL.124u1102A} and CREAM
\cite{2017ApJ...839....5Y}, with the background plus nearby
source model. In each panel, the blue line is the background
component, the red is the nearby source component, and the
black is their sum.
\label{fig:spec}}
\end{figure*}

Fig.~\ref{fig:spec} displays the comparison between the model predictions
and the measurements, for the energy spectra of He, C, O, and Ne, Mg, Si
species of CRs. In this calculation, the spectral indices of the background
are 2.20 and 2.36 for rigidities below and above 7.2 GV, and the cutoff 
rigidity is about 7.0 PV. The spectral index for the nearby source is 
2.06, and the cutoff rigidity is about 30 TV. 
Note that the model parameters differ slightly from that of previous
works \cite{2019JCAP...10..010L,2019JCAP...12..007Q}, due partly to the
inclusion of new AMS-02 Ne, Mg, Si data in the fitting. Furthermore, we 
extend the fitting to energies below 100 GeV, taking into account the 
solar modulation effect, which is expected to be more self-consistent.
The nearby source is assumed to be located at $l=170^{\circ}$, 
$b=-20^{\circ}$. Its distance is adopted to be $\sim0.33$~kpc, and its 
age is $3.4\times10^5$ yr, which are similar with that of (the birth 
place of) Geminga \cite{1994A&A...281L..41S,2005AJ....129.1993M}. 
As for the relative abundances, the ratio of the nearby source to the 
background is assumed to be 1.7 times lower for the Ne, Mg, Si group than 
that for the He, C, O group. To fit the low-energy data, a force-field 
solar modulation model with a modulation potential of $0.4$~GV is applied 
\cite{1968ApJ...154.1011G}. 
It is shown that this model can well describe the data.

For the background spectra, the spatially-dependent diffusion can
give a gradual spectral hardening, due to the fact that the
rigidity-dependence slope of the diffusion coefficient is smaller
in the disk, resulting in a harder high-energy component
\cite{2016PhRvD..94l3007F}. This property should be universal for
all species, and thus is not enough to account for the differences
between the He, C, O group and the Ne, Mg, Si group. As we have
discussed before, the spatially-dependent diffusion is well motivated
by the $\gamma$-ray and the very-high-energy anisotropy observations,
which is therefore included in this work.

\begin{figure*}[!htb]
\centering
\includegraphics[width=\textwidth]{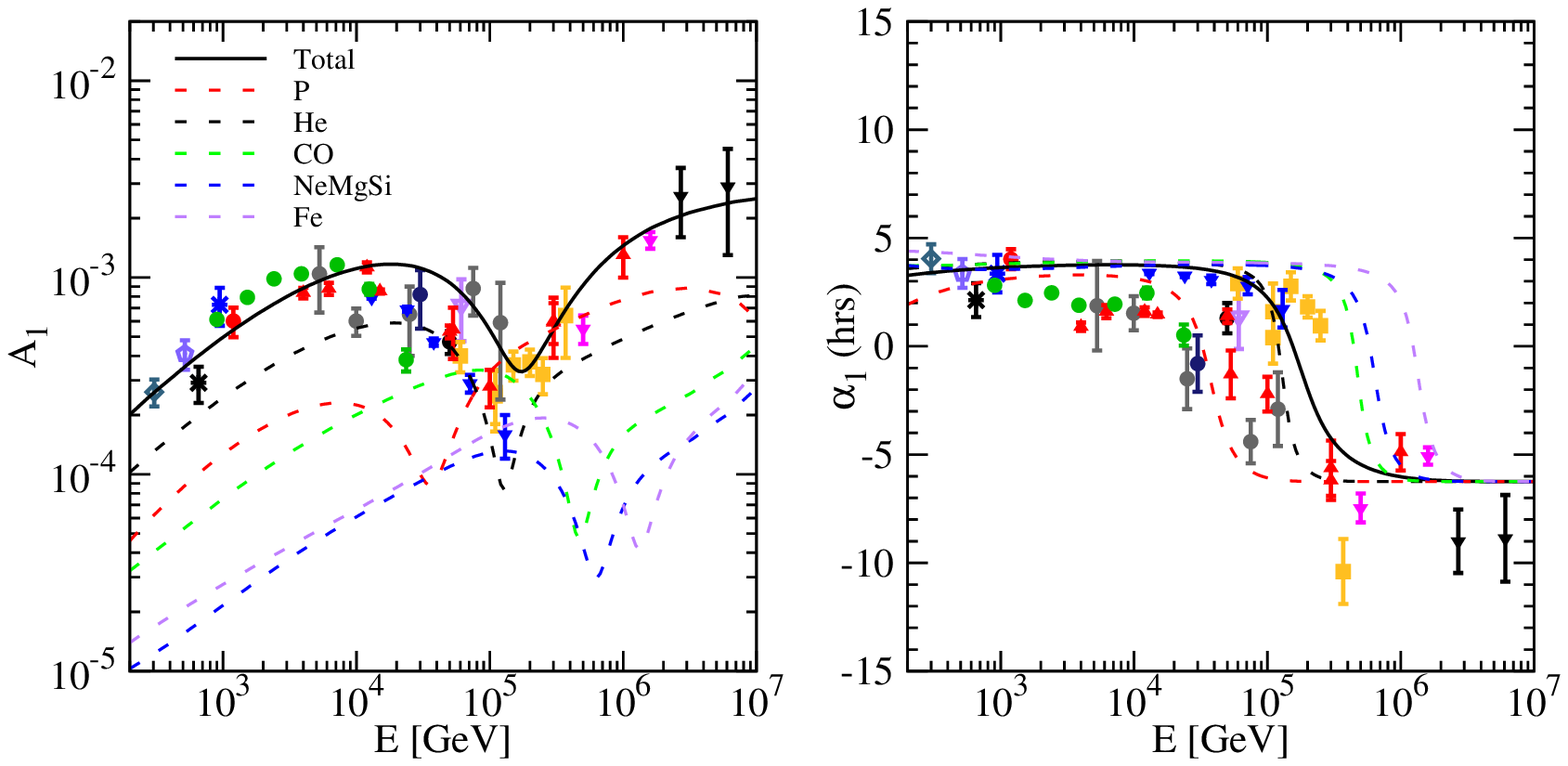}
\caption{The energy dependences of the amplitudes (left) and phases (right)
of the dipole anisotropies. All the major CR species have been included.
The data are from \cite{1973ICRC....2.1058S,1981ICRC...10..246B,
1981ICRC....2..146A,2009NuPhS.196..179A,1987ICRC....2...22A,
1985P&SS...33.1069S,1995ICRC....4..639M,1995ICRC....4..648M,
1995ICRC....4..635F,2003PhRvD..67d2002A,1975ICRC....2..586G,
1995ICRC....2..800A,1996ApJ...470..501A,2009ApJ...692L.130A,
2015ICRC...34..281C,2010ApJ...718L.194A,2013ApJ...765...55A,
2015ApJ...809...90B,2005ApJ...626L..29A,2015ICRC...34..355A,
2017ApJ...836..153A}.
}
\label{fig:aniAll}
\end{figure*}

\begin{figure*}[!htb]
\centering
\includegraphics[width=\textwidth]{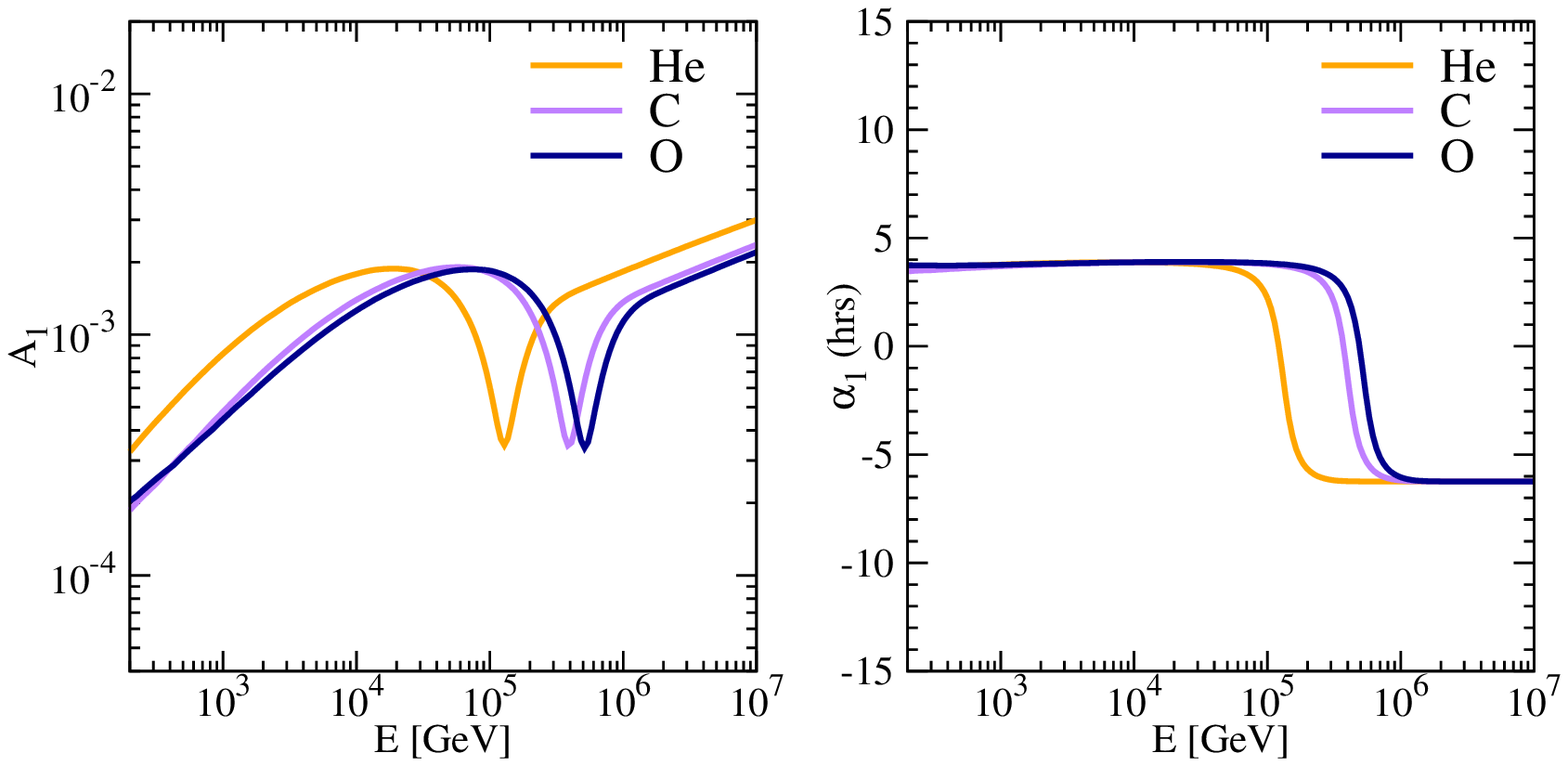}
\includegraphics[width=\textwidth]{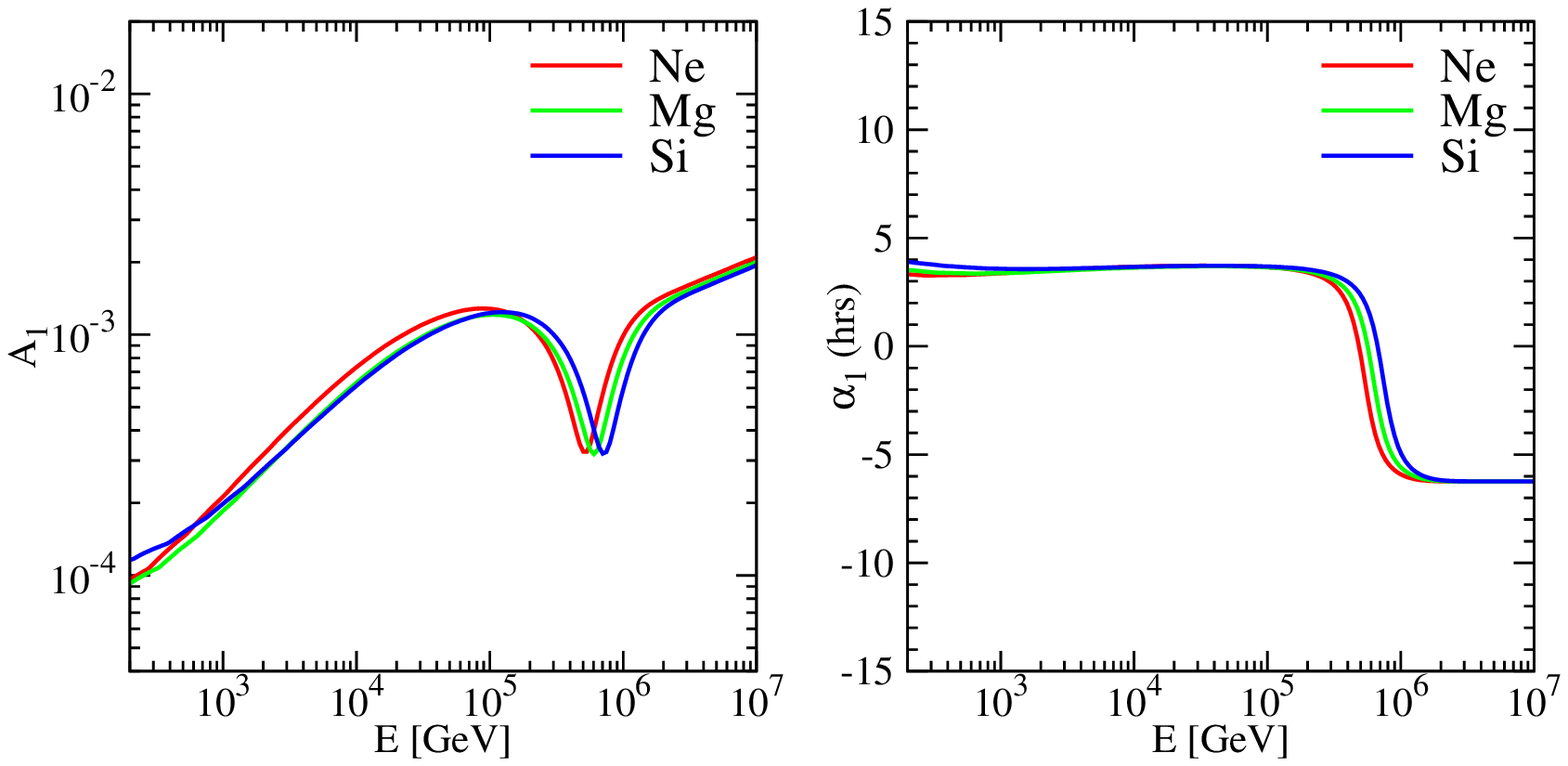}
\caption{The energy dependence of the amplitudes (left) and phases (right)
of the dipole anisotropies when adding all of the major elements together.
The data points are taken from underground muon detectors:
}
\label{fig:aniZ}
\end{figure*}

Fig.~\ref{fig:aniAll} shows the model predicted amplitudes and phases 
of the dipole anisotropies as functions of energies, compared with
the data. The dip of the amplitudes and phase-flipping around 100
TeV are due to the transition of the dominant component of the CR
streamings from the nearby source to the background component, as 
shown in Ref.~\cite{2019JCAP...10..010L}. The contribution to the
energy spectra from the nearby source is, however, sub-dominant
compared with the background component. 
It is interesting to note that below the dip, the anisitropies
are dominated by the helium component. In this model, the anisotropy 
amplitude is sensitive to the relative flux differences between the
background component and the nearby source component. Due to a 
relatively high helium contribution from the nearby source
\cite{2019JCAP...12..007Q}, helium nuclei dominate the total
anisotropies of all CR particles in the low energy range.

The differences of the element abundances between the nearby source
and the background directly imprint on the anisotropies of different
species, as shown in Fig.~\ref{fig:aniZ}. The peak values of the
anisotropy amplitudes around 100 TeV, which are mainly due to the
nearby source, show a difference of $\sim1.5$ between the He, C, O
group and the Ne, Mg, Si group. The forthcoming measurements of the
evolution of anisotropies of different mass groups by e.g., the
Large High Altitude Air Shower Observatory \cite{2019arXiv190502773B}
may test this prediction.

\section{Conclusion and discussion}

In this work we employ the nearby source scenario to explain the
newest measurements of spectral structures of CRs. This simple model
can naturally explain the spectral hardenings of CR nuclei around
200 GV, the softenings around 10 TV, and the energy-dependence of
the amplitudes and phases of the large-scale anisotropies.
The observed spectral differences between the He, C, O group and
the Ne, Mg, Si group can be understood as the slightly different
element abundances of the nearby source from that of the background
sources. It is natural that the source abundances of CRs differ
from one to another, depending on e.g., the progenitor star's
properties and/or the environments of the CR acceleration.
The amplitudes of low-energy ($<$PeV) anisotropies, which are 
dominated by the nearby source in this model, are smaller by
a factor of $\sim1.5$ for the Ne, Mg, Si group than the He, C, O 
group. This prediction can be tested with future measurements
of anisotropies of different mass groups.

To fit the data, it is required that the nearby source has relatively
higher abundances of He, C, O, compared with Ne, Mg, Si. There are
many factors affecting the nucleosynthesis inside a star. Some key
parameters include the mass, initial metalicity, rotation, convection,
and so on. It is likely that a star with relatively higher mass or
higher spin tends to generate less Ne, Mg, Si, compared with a lower 
mass/spin star (e.g., \cite{2005A&A...433.1013H}). Therefore, the 
AMS-02 results may suggest that the progenitor of the nearby source 
is a relatively high-mass/high-spin star.

It is also possible that the acceleration of different elements 
at the source may give such a difference. The particle acceleration
depends on the shock properties and the environment parameters.
Although all these species discussed in this work have $A/Z\approx2$,
their ionization histories may be different due to different energy
levels of electrons. The ionization histories may affect the 
injection and acceleration efficiency of the nuclei, resulting in
different abundances in CRs (e.g., \cite{1978ApJ...221..703C}).
Alternatively, it was expected that the condensation of elements
into grains affect the acceleration efficiencies of different
species (e.g., \cite{1997ApJ...487..197E}). The so-called
refractory elements such as Mg, Al, Si are likely locked into 
grains are accelerated more efficiently than in the interstellar
gas phase. If the dust fraction of the nearby source environment is 
smaller than that of the Milky Way average, the relative abundances 
of the Ne, Mg, Si particles could be lower.

\acknowledgments
We thank Hai-Ning Li, Xin-Lian Luo, and Bo Zhang for helpful discussion.
This work is supported by the National Key Research and Development
Program of China (Nos. 2018YFA0404203, 2016YFA0400200), the National 
Natural Science Foundation of China (Nos. 11722328, 11525313, U1738205, 
11851305), the 100 Talents program of Chinese Academy of Sciences, 
and the Program for Innovative Talents and Entrepreneur in Jiangsu.

\end{document}